\documentstyle[12pt,epsf]{article}
\def\slash#1{/ \hskip -0.5em #1}
\def\vq1{\slash{q}_1}
\def\bok{\slash{k}}

\catcode`\@=11
\long\def\@makefntext#1{
\protect\noindent \hbox to 3.2pt {\hskip-.9pt
$^{{\ninerm\@thefnmark}}$\hfil}#1\hfill}                

\def\@makefnmark{\hbox to 0pt{$^{\@thefnmark}$\hss}}  

\def\ps@myheadings{\let\@mkboth\@gobbletwo
\def\@oddhead{\hbox{}
\rightmark\hfil\ninerm\thepage}
\def\@oddfoot{}\def\@evenhead{\ninerm\thepage\hfil
\leftmark\hbox{}}\def\@evenfoot{}
\def\sectionmark##1{}\def\subsectionmark##1{}}

\renewcommand{\thefootnote}{\fnsymbol{footnote}}

\def\sectionc{\@startsection {section}{1}{\z@}{-3.5ex plus -1ex minus
    -.2ex}{2.3ex plus .2ex}{\bf }}
\def\subsectionc{\@startsection{subsection}{2}{\z@}{-3.25ex plus -1ex minus
   -.2ex}{1.5ex plus .2ex}{\it }}
\renewcommand{\section}[1]{\sectionc{#1}\hspace*{\parindent}}
\renewcommand{\subsection}[1]{\subsectionc{#1}\hspace*{\parindent}}
\newcounter{appendixc}
\newcounter{subappendixc}[appendixc]
\newcounter{subsubappendixc}[subappendixc]

\renewcommand{\appendix}[1] {\vspace*{0.6cm}
        \refstepcounter{appendixc}
        \setcounter{figure}{0}
        \setcounter{table}{0}
        \setcounter{equation}{0}
        \renewcommand{\thefigure}{\Alph{appendixc}.\arabic{figure}}
        \renewcommand{\thetable}{\Alph{appendixc}.\arabic{table}}
        \renewcommand{\theappendixc}{\Alph{appendixc}}
        \renewcommand{\theequation}{\Alph{appendixc}.\arabic{equation}}
        \noindent{\bf Appendix \theappendixc #1}\par\vspace*{0.4cm}}

\def\abstracts#1{{

\centering{\begin{minipage}{13.2truecm}\footnotesize\baselineskip=13pt
\noindent
        \parindent=0pt #1
        \end{minipage}}\par}}

\renewenvironment{thebibliography}[1]
        {\begin{list}{\arabic{enumi}.}
        {\usecounter{enumi}\setlength{\parsep}{0pt}
\setlength{\leftmargin 0.75cm}{\rightmargin 0pt}
         \setlength{\itemsep}{0pt} \settowidth
        {\labelwidth}{#1.}\sloppy}}{\end{list}}

\newcounter{itemlistc}
\newcounter{romanlistc}
\newcounter{alphlistc}
\newcounter{arabiclistc}

\newcommand{\fcaption}[1]{
        \refstepcounter{figure}
        \setbox\@tempboxa = \hbox{\footnotesize Figure~\thefigure. #1}
        \ifdim \wd\@tempboxa > 6in
           {\begin{center}
        \parbox{6in}{\footnotesize\baselineskip=13pt Figure~\thefigure. #1}
            \end{center}}
        \else
             {\begin{center}
             {\footnotesize Figure~\thefigure. #1}
              \end{center}}
        \fi}

\newcommand{\tcaption}[1]{
        \refstepcounter{table}
        \setbox\@tempboxa = \hbox{\footnotesize Table~\thetable. #1}
        \ifdim \wd\@tempboxa > 6in
           {\begin{center}
        \parbox{6in}{\footnotesize\baselineskip=13pt Table~\thetable. #1}
            \end{center}}
        \else
             {\begin{center}
             {\footnotesize Table~\thetable. #1}
              \end{center}}
        \fi}

\def\@citex[#1]#2{\if@filesw\immediate\write\@auxout
        {\string\citation{#2}}\fi
\def\@citea{}\@cite{\@for\@citeb:=#2\do
        {\@citea\def\@citea{,}\@ifundefined
        {b@\@citeb}{{\bf ?}\@warning
        {Citation `\@citeb' on page \thepage \space undefined}}
        {\csname b@\@citeb\endcsname}}}{#1}}

\newif\if@cghi
\def\cite{\@cghitrue\@ifnextchar [{\@tempswatrue
        \@citex}{\@tempswafalse\@citex[]}}
\def\citelow{\@cghifalse\@ifnextchar [{\@tempswatrue
        \@citex}{\@tempswafalse\@citex[]}}
\def\@cite#1#2{{$\null^{#1}$\if@tempswa\typeout
        {IJCGA warning: optional citation argument
        ignored: `#2'} \fi}}

 1
 1
 1

\font\ninerm=cmr9

\begin{document}

\centerline{\normalsize\bf A MODEL FOR TWO-PION PHOTOPRODUCTION AMPLITUDES 
\footnote{
Supported by DOE contracts DE-AC05-84ER40150 and DE
FG05-94ER40832, NSF award PHY 9457892, and by the Thomas F. Jeffress and Kate 
Miller 
Jeffress Memorial Trust.}}
\baselineskip=15pt

\vspace*{0.6cm}
\centerline{\footnotesize W. Roberts}
\baselineskip=13pt
\centerline{\footnotesize\it Department of Physics, Old Dominion University}
\baselineskip=13pt
\centerline{\footnotesize\it Norfolk, VA 23529, USA,}
\baselineskip=13pt
\centerline{\footnotesize\it Thomas Jefferson National Accelerator Facility}
\baselineskip=13pt
\centerline{\footnotesize\it12000 Jefferson Ave., Newport News, VA 23606}
\baselineskip=13pt
\centerline{\footnotesize E-mail: roberts@jlab.org}
\vspace*{0.3cm}
\centerline{\footnotesize and}
\vspace*{0.3cm}
\centerline{\footnotesize A. Rakotovao}
\baselineskip=13pt
\centerline{\footnotesize\it Department of Physics, Old Dominion University}
\baselineskip=13pt
\centerline{\footnotesize\it Norfolk, VA 23529, USA}
\baselineskip=13pt
\centerline{\footnotesize E-mail: rakotova@jlab.org}

\vspace*{0.6cm}
\abstracts{We present a brief discussion of the general form of the amplitude 
that 
describes the
two-pion photoproduction process. We outline an effective Lagrangian method 
that we 
are using to
calculate this amplitude, and comment briefly on a few aspects of the 
calculation.}

\normalsize\baselineskip=15pt
\setcounter{footnote}{0}
\renewcommand{\thefootnote}{\alph{footnote}}

\section{Introduction}

Experiments in which two pions are photoproduced are expected to play a major
role in our quest for the baryons that have been predicted to exist in various
versions of the quark model, but for which there is little evidence in the
present analyses of the experimental data \cite{scwr}. The high precisions 
expected 
from
Jefferson Laboratory, and from other facilities around the globe, as well as
the various polarization measurements, should allow
us to probe all but the very weakest of these apparently weakly coupled states.
On the other hand, a systematic analysis of two-pion photoproduction processes
that is similar to those which exist for single pion photoproduction, has not,
as far as we know, yet been attempted.

In the past, this process has been analyzed by assuming that the final state
arises mainly from a quasi-two-body process such as $\gamma N\to N\rho\to 
N\pi\pi$ 
or
$\gamma N\to\Delta\pi\to N\pi\pi$ \cite{luke}. Application of kinematic cuts 
consistent
with this assumption then yielded some set of results. The recent calculation 
of
Oset {\it et al.} \cite{oset} should highlight the dangers of such a procedure, 
however. In
their calculation, they attempt to fit the recent Mainz data \cite{mainz} by 
including a number
of resonant and non-resonant terms in an effective Lagrangian approach. They 
find that
the dominant contribution to the cross section comes not from either of the two
processes described above, but from the interference of one `$\Delta\pi$'
resonant contribution with one of their non-resonant terms. In this case, 
performing
kinematic cuts on the data for the purpose of analyzing them in terms of the
quasi-two-body channels $\Delta\pi$ and $N\rho$ would yield meaningless
results. It is also not at all intuitive that such an interference term should
dominate the cross section, in {\it any} energy range.

With this said, it should be quite clear that a systematic theoretical
treatment of the two-pion photoproduction (and electroproduction) process is
essential as the starting point for a comprehensive analysis of the high
quality data expected. In addition, we believe that it is crucial that future
analyses move away from `parametrizing the non-resonant background' to
`understanding the apparently non-resonant background', as valuable information
may be hidden in this background.

To this end, we have undertaken to construct a model for two-pion
photoproduction that is as general as possible. An outline of our procedure is
the subject of this article: we can not attempt to describe all of the details
of the calculation in these few pages, nor can we present any final results, as
this is very much work in progress.

In the next section of this note, after a brief discussion of the kinematics of
the reaction, we describe the most general form that the amplitude for this
process must take. This amplitude requires a certain number of form factors, and
we next describe how we calculate the contributions to these form factors,
using the effective Lagrangian approach. After a brief discussion of the merits
and disadvantages of this approach, we conclude with some comments on how our
analysis may lead to new information on the structure of hadrons, both baryons
and mesons.

\section{Amplitude}

\subsection{Kinematics}

\begin{figure}
\centerline{\epsfysize=4in \epsfbox{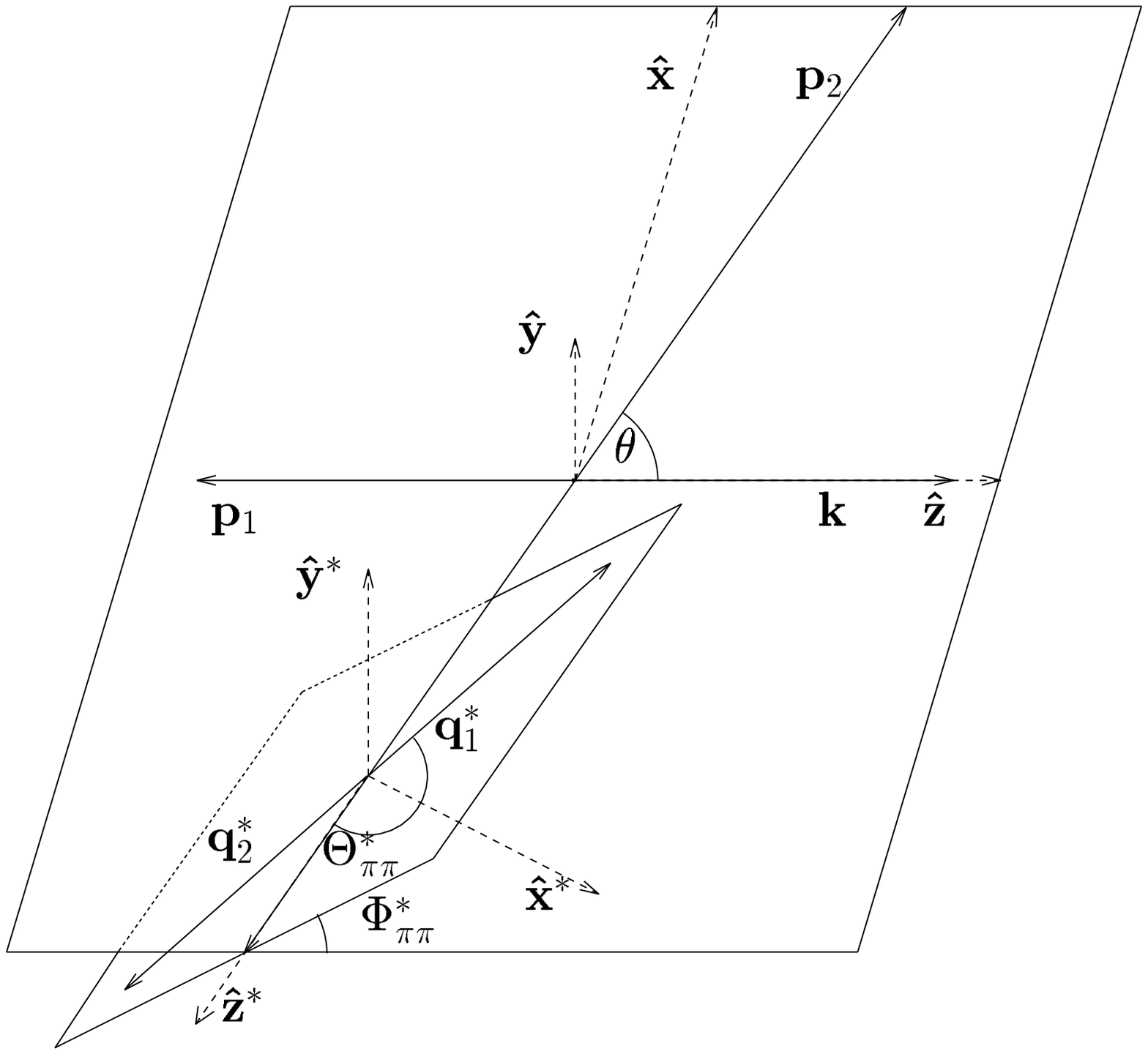}}
\vspace*{-1.5in}
\fcaption{Kinematics of the two-pion photoproduction process. In the figure, 
the 
axes $\hat{y}=
\hat{y}^*$, $\hat{z}^*=\frac{\vec{q}_1+\vec{q}_2}{\left|\vec{q}_1+
\vec{q}_2\right|}$, and
$\hat{x}^*=\hat{z}^*\times\hat{y}^*$.\label{fig:fig1}}
\end{figure}

Before we present the amplitude for the process of interest, we first describe 
the 
kinematics of the
process. The kinematics are shown schematically in figure \ref{fig:fig1}. $k$ 
is 
the momentum of the photon,
$p_1$ is that of the target nucleon, $p_2$ is that of the scattered nucleon, 
and 
$q_1$ and $q_2$ are the
pion momenta. Momentum conservation gives
\begin{equation}
k+p_1=p_2+q_1+q_2.
\end{equation}
Thus, when we construct the amplitude for the process using all the 
four-vectors 
at our disposal, we can
eliminate one of these from consideration. 

The total center-of-mass energy of the process is $\sqrt{s}$, where
$s=(k+p_1)^2$. We may define a variable $t$ as the square of the momentum
transfered to the nucleon. Thus, $t=(p_2-p_1)^2$, which can be related to the
scattering angle of the nucleon in the c-o-m frame.

The differential cross section for this process is described in terms of 5
kinematic variables. These may be, for instance, two Lorentz invariants and
three angles. One obvious choice for one of the invariants is $s$. The choice
of the other four quantities can be fairly arbitrary, and will depend on what
information is being presented. One choice is the scattering angle of the 
nucleon,
$\theta$, or equivalently, $t$. For the other three variables, we can choose,
for example, $s_{\pi\pi}\equiv (q_1+q_2)^2$ and $d\Omega_{\pi\pi}^*\equiv
d\Theta_{\pi\pi}^*d\Phi_{\pi\pi}^*$, as illustrated in the figure. Another
equally valid choice would be $s_{N\pi_1}\equiv(p_2+q_1)^2$ and 
$d\Omega_{N\pi_1}^*$, where the solid angle is defined in the rest frame of the
nucleon-pion pair.

\subsection{General Amplitude}

Our starting point is the identification of the most general form for the
transition amplitude for this process. While the requirements of Lorentz
covariance and gauge invariance delimit the form of the amplitude, we find that
there is nevertheless quite a bit of freedom in the form chosen. The most 
general 
form is 
\begin{equation}
i {\cal M}=\overline{U_f(p_2)}\varepsilon_{\mu} {\cal O}^\mu U_f(p_1)
\end{equation}
where
\begin{eqnarray} \label{eq:Omu}
{\cal O}^\mu &=& a_1 p_1^\mu + a_2 p_2^\mu + a_3 q_1^\mu + a_4 \gamma^\mu
+\slash{k}\left(a_5 p_1^\mu + a_6 p_2^\mu + a_7 q_1^\mu + a_8 \gamma^\mu\right)
\nonumber \\ 
&+& \slash{q}_1 \left(a_9 p_1^\mu + a_{10} p_2^\mu +a_{11}q_1^\mu +a_{12} 
\gamma^\mu\right)
+\slash{q}_1 \slash{k}\left(a_{13} p_1^\mu + a_{14} p_2^\mu +a_{15}q_1^\mu+
a_{16} 
\gamma^\mu \right).
\end{eqnarray}
Note that we have no terms in $\slash{p}_1$ nor $\slash{p}_2$, as the initial 
and 
final nucleons each
satisfy
\begin{equation}
\slash{p}U(p)=m U(p).
\end{equation}

The form factors $a_i$ are all functions of the kinematic variables $s$, 
$s_{\pi\pi}$, $\theta$, $\theta^*$ and $\phi^*$, or whatever combination of 
kinematic variables is
chosen. Their exact dependence on each of these variables will
be determined by the specific model constructed.

Gauge invariance of the amplitude requires that $k_\mu {\cal O}^\mu=0$, which 
leads to the four relations
\begin{eqnarray}
&&a_1k\cdot p_1 +a_2 k\cdot p_2 +a_3 q_1\cdot k=0,\\
&&a_4+a_5k\cdot p_1 +a_6 k\cdot p_2 +a_7 q_1\cdot k=0,\\
&&a_9k\cdot p_1 +a_{10} k\cdot p_2 +a_{11} q_1\cdot k=0,\\
&&a_{12}+a_{13}k\cdot p_1 +a_{14} k\cdot p_2 +a_{15} q_1\cdot k=0.
\end{eqnarray}
Note that there is no condition on either of the form factors $a_8$ or 
$a_{16}$.

From these equations, we can eliminate four of the form factors, leaving us 
with twelve independent form
factors, or Lorentz-Dirac structures, to describe the amplitude. One choice 
would be to eliminate $a_1$, $a_4$, $a_9$, $a_{12}$, giving
\begin{eqnarray}
\varepsilon_\mu {\cal O^\mu} &=&                                                        
\left\{ \frac{1}{p_1\cdot k}\left[(a_2 + a_{10} \vq1)p_{2\mu} p_{1\nu} + 
(a_3 + a_{11} \vq1) 
q_{1\mu} p_{1\nu}\vphantom{\frac{1}{2}}\right]+(a_5 + a_{13} \vq1)p_{1\nu} 
\gamma_\mu \right. \nonumber \\
&+& \left.(a_6 + a_{14} \vq1) p_{2\mu} \gamma_\nu +(a_7 + a_{15} \vq1) 
q_{1\mu} 
\gamma_\nu \vphantom{\frac{1}{p_1\cdot k}}-\frac{1}{2}                   
(a_8 + a_{16} \vq1) \gamma_\mu \gamma_\nu\right\}F^{\mu\nu}, \nonumber  
\end{eqnarray}
where $F^{\mu\nu}=\varepsilon^\mu k^\nu-\varepsilon^\nu k^\mu$.
Another choice is $a_1$, $a_5$, $a_9$, $a_{13}$, giving
\begin{eqnarray}
\varepsilon_\mu {\cal O^\mu} &=&                                                        
\left\{ \frac{1}{p_1\cdot k}\left[\vphantom{\frac{1}{p_1\cdot k}}\left(a_2 +
\bok a_6+ \vq1 a_{10} +\vq1\bok a_{14}\right)p_{2\mu} p_{1\nu}
+\left(a_4 + \vq1 a_{12}\right)p_{1\nu} \gamma_\mu\right.\right.\nonumber\\                       
&+&\left.\left.\left(a_3 + \bok a_7 + \vq1 a_{11}+ \vq1\bok a_{15}\right)
q_{1\mu} p_{1\nu}
\vphantom{\frac{1}{p_1\cdot k}}\right]
-\frac{1}{2}\left(a_8 + a_{16} \vq1\right) \gamma_\mu \gamma_\nu\right\}
F^{\mu\nu}. \nonumber  
\end{eqnarray}

In terms of the $a_i$ we can generate CGLN-type \cite{cgln} amplitudes from 
the structures given. We write the amplitude as \cite{toed}
\begin{equation}
i {\cal M}=\psi_f^\dagger {\cal F} \psi_i,
\end{equation}
with
\begin{eqnarray}
{\cal F}&=&
\vec{\varepsilon} \cdot \vec{q_1} F_1+\vec{\varepsilon} \cdot \vec{p_2} F_2 
+\vec{\varepsilon} \cdot \vec{q_1} \vec{\sigma} \cdot \vec{p_2} \vec{\sigma} 
\cdot \vec{q_1} F_3
+\vec{\varepsilon} \cdot \vec{p_2} \vec{\sigma} \cdot \vec{p_2} \vec{\sigma} 
\cdot \vec{q_1} F_4\nonumber\\ 
&+&\vec{\varepsilon} \cdot \vec{q_1} \vec{\sigma} \cdot \vec{p_2} \vec{\sigma} 
\cdot \vec{k} F_5
+\vec{\varepsilon} \cdot \vec{p_2} \vec{\sigma} \cdot \vec{p_2} \vec{\sigma} 
\cdot \vec{k} F_6
+\vec{\varepsilon} \cdot \vec{q_1} \vec{\sigma} \cdot \vec{q_1} \vec{\sigma} 
\cdot 
\vec{k} F_7
+\vec{\varepsilon} \cdot \vec{p_2} \vec{\sigma} \cdot \vec{q_1} \vec{\sigma} 
\cdot 
\vec{k} F_8
\nonumber\\ 
&+&\vec{\sigma} \cdot \vec{q_1} \vec{\sigma} \cdot \vec{\varepsilon} F_9
+\vec{\sigma} \cdot \vec{p_2} \vec{\sigma} \cdot \vec{\varepsilon} F_{10} 
+\vec{\sigma} \cdot (\vec{\varepsilon} \times \vec{k}) F_{11}
+\vec{\sigma} \cdot \vec{p_2} \vec{\sigma} \cdot \vec{q_1} \vec{\sigma} \cdot 
(\vec{\varepsilon} \times \vec{k}) F_{12},
\end{eqnarray}
or
\begin{eqnarray}
{\cal F}&=&\vec{\varepsilon} \cdot \vec{q_1} F^\prime_1
+\vec{\varepsilon} \cdot \vec{p_2} F^\prime_2 
+\vec{\varepsilon} \cdot \vec{q_1} \vec{\sigma} \cdot \vec{p_2} \times 
\vec{q_1} F^\prime_3
+\vec{\varepsilon} \cdot \vec{p_2} \vec{\sigma} \cdot \vec{p_2} \times 
\vec{q_1} F^\prime_4 \nonumber\\ 
&+&\vec{\varepsilon} \cdot \vec{q_1} \vec{\sigma} \cdot \vec{p_2} \times 
\vec{k} F^\prime_5
+\vec{\varepsilon} \cdot \vec{p_2} \vec{\sigma} \cdot \vec{p_2} \times \vec{k} 
F^\prime_6
+\vec{\varepsilon} \cdot \vec{q_1} \vec{\sigma} \cdot \vec{q_1} \times \vec{k} 
F^\prime_7
+\vec{\varepsilon} \cdot \vec{p_2} \vec{\sigma} \cdot \vec{q_1} \times \vec{k} 
F^\prime_8
\nonumber\\ 
&+&\vec{\sigma} \cdot \vec{q_1} \times \vec{\varepsilon} F^\prime_9
+\vec{\sigma} \cdot \vec{p_2} \times \vec{\varepsilon} F^\prime_{10} 
+\vec{\sigma} \cdot (\vec{\varepsilon} \times \vec{k}) F^\prime_{11}
+\vec{\sigma} \cdot \vec{p_2} \times \vec{q_1} \vec{\sigma} \cdot 
(\vec{\varepsilon} \times \vec{k}) 
F^\prime_{12}.
\end{eqnarray}
Neither of these may be the optimal forms, as we are yet to explore which 
choice of 
structures will lead to the
simplest representation of helicity amplitudes, differential cross sections, 
etc. 
In these last two
expressions, the $F_i$ and $F_i^\prime$ are linear combinations of the $a_i$.

We close this section with a short comment. The form we have written is the 
most 
general that can be written
for this amplitude. All quantities of interest can now be calculated in terms 
of 
the form factors, the $a_i$.
Thus, up to this point, everything has been completely model independent. 
All that 
is now left to be done is to
construct a model for the $a_i$. A variety of approaches are possible, but we 
confine ourselves to a brief
discussion of the model we have chosen for this work.

\section{Effective Lagrangian}

The approach we use to calculate the $a_i$ introduced in the last section is 
that 
of the
effective Lagrangian. In this approach, we treat all particles as essentially 
point-like;
any structure in these particles will be accounted for by the introduction of 
appropriate
form factors. Next, a set of vertices involving these point-like particles is 
defined, and
a set of `Feynman' diagrams describing the process of interest is drawn. From 
each of these
diagrams, the contribution to each of the $a_i$ is extracted, and the $a_i$ are 
thus built
from a number of such diagrams. We note that we perform the calculation only at 
tree level,
as it would become rapidly intractable if loops are included, as will be 
apparent 
from what follows.

To date, in our calculation, we have included nucleons and $\Delta$'s of spin 
1/2 
and 3/2,
as well as vector mesons and, of course, pions. As it is our aim to make our 
calculation as
general as possible from the outset, we do not identify, for instance, any 
particular spin
1/2 resonance, but simply include a generic set of terms that would be valid 
for 
any spin 1/2
resonance.

For spin 1/2 resonances, as well as for the ground state nucleon, the wave 
function 
is a
Dirac spinor $u(p)$ satisfying
\begin{equation}
\slash{p}u(p)=mu(p),
\end{equation}
where $m$ is the mass of the resonance. The corresponding propagator is
\begin{equation}
\frac{\slash{p}+m}{p^2-m^2+im\Gamma},
\end{equation}
where $\Gamma$ is the total width of the state. For the ground-state nucleon, 
the
$+im\Gamma$ in the propagator is replaced by the usual $+i\epsilon$.

In the case of the spin-3/2 baryons, the wave function is the Rarita-Schwinger 
field $u_\mu$, with
\begin{equation}
\slash{p}u_\mu(p)=mu_\mu(p).
\end{equation}
$u_\mu$ also satifies the auxiliary conditions
\begin{equation}
p_\mu u^\mu(p)=0,\,\,\,\,\gamma_\mu u^\mu(p)=0,
\end{equation}
and the appropriate propagator is
\begin{equation}
\Theta_{\mu\nu}=\frac{\slash{p}+m}{p^2-m^2+im\Gamma}\left(g_{\mu\nu}-
\frac{1}{3}
\gamma_\mu\gamma_\nu+\frac{p_\mu\gamma_\nu-p_\nu\gamma_\mu}{3m}-
\frac{2p_\mu p_\nu}{3m^2}\right).
\end{equation}
In writing this form for the propagator, we are ignoring the so-called 
off-shell
contributions, at least for the time being. The off-shell contributions may 
also 
be included at the vertices,
as has been done, for example, by Benmerrouche {\it et. al} \cite{nimai}.

In addition to the pions and the photon, the treatment of which is presumed to 
be 
well
known, the only other particles that we have in our calculation to date are 
the 
vector
mesons, which are described by the polarization vector $\varepsilon_\mu$. This 
satisfies
\begin{equation}
\varepsilon_\mu(p) p^\mu=0,
\end{equation}
and the corresponding propagator is
\begin{equation}
\Theta^{\mu\nu}=\frac{g^{\mu\nu}-\frac{p^\mu p^\nu}{p^2}}{p^2-m^2+im\Gamma}.
\end{equation}

\begin{figure}
\centerline{\epsfysize=5in \epsfbox{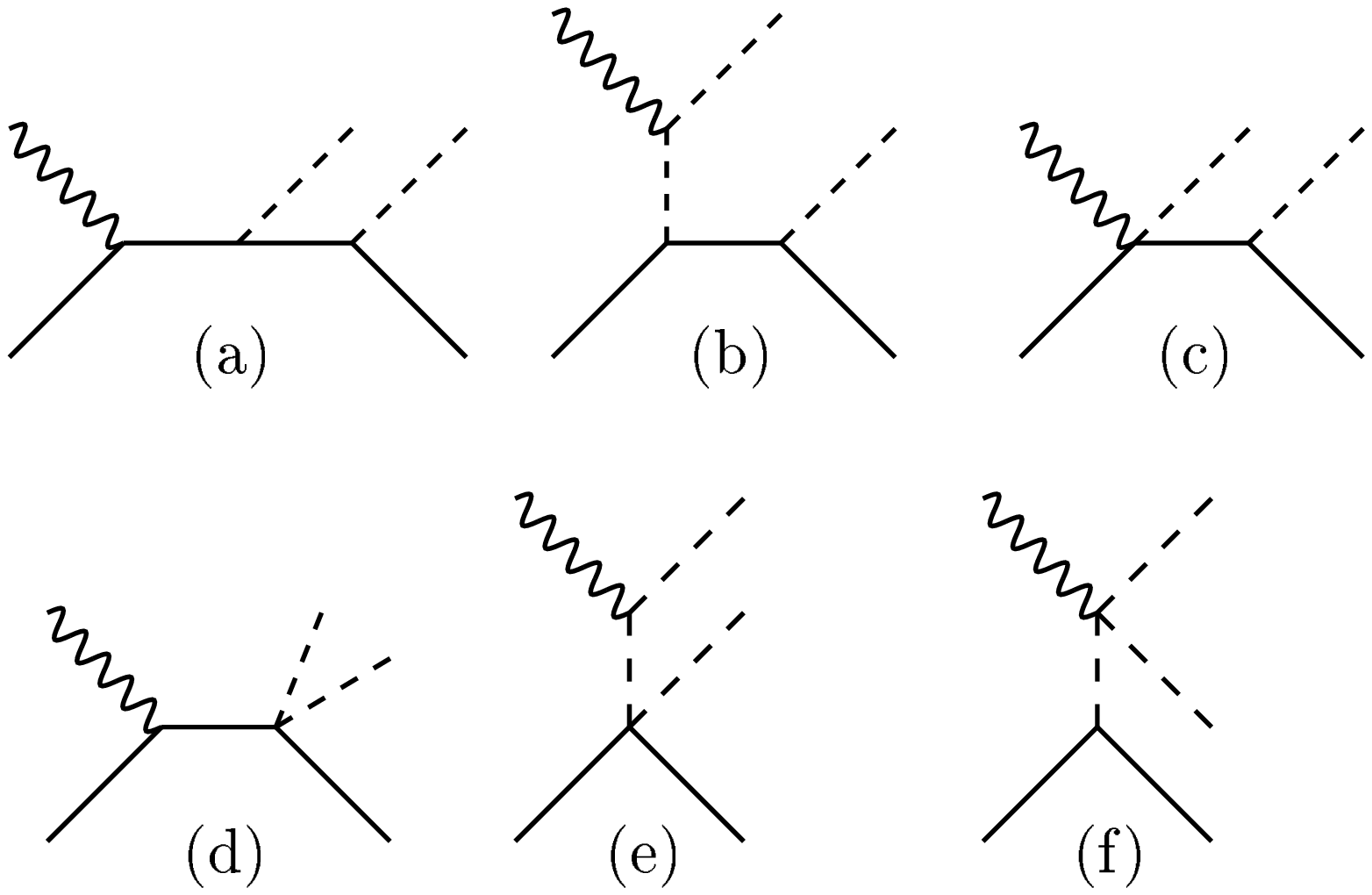}}
\vspace*{-3in}
\fcaption{Born diagrams: diagrams containing only ground-state nucleons and
pions.\label{fig:born}}
\end{figure}

Once these fields have been defined, we next proceed to construct effective 
Lagrangians for
them, concentrating on the interaction vertices. With these vertices, one can 
construct 
a number of diagrams that contribute to the amplitude, and hence to each of the 
$a_i$. Space does not permit us to show all of the vertices that enter into
this calculation, so we spare the reader the torture of several tens of vertex 
diagrams. Instead, we show some of
the diagrams that we draw for the process of interest. In all of the diagrams, 
solid lines represent the ground
state nucleons, thick solid lines are baryon resonances, wavy lines are 
photons, 
dashed lines are pions, and curly
lines are vector mesons. 

\begin{figure}
\centerline{\epsfysize=5in \epsfbox{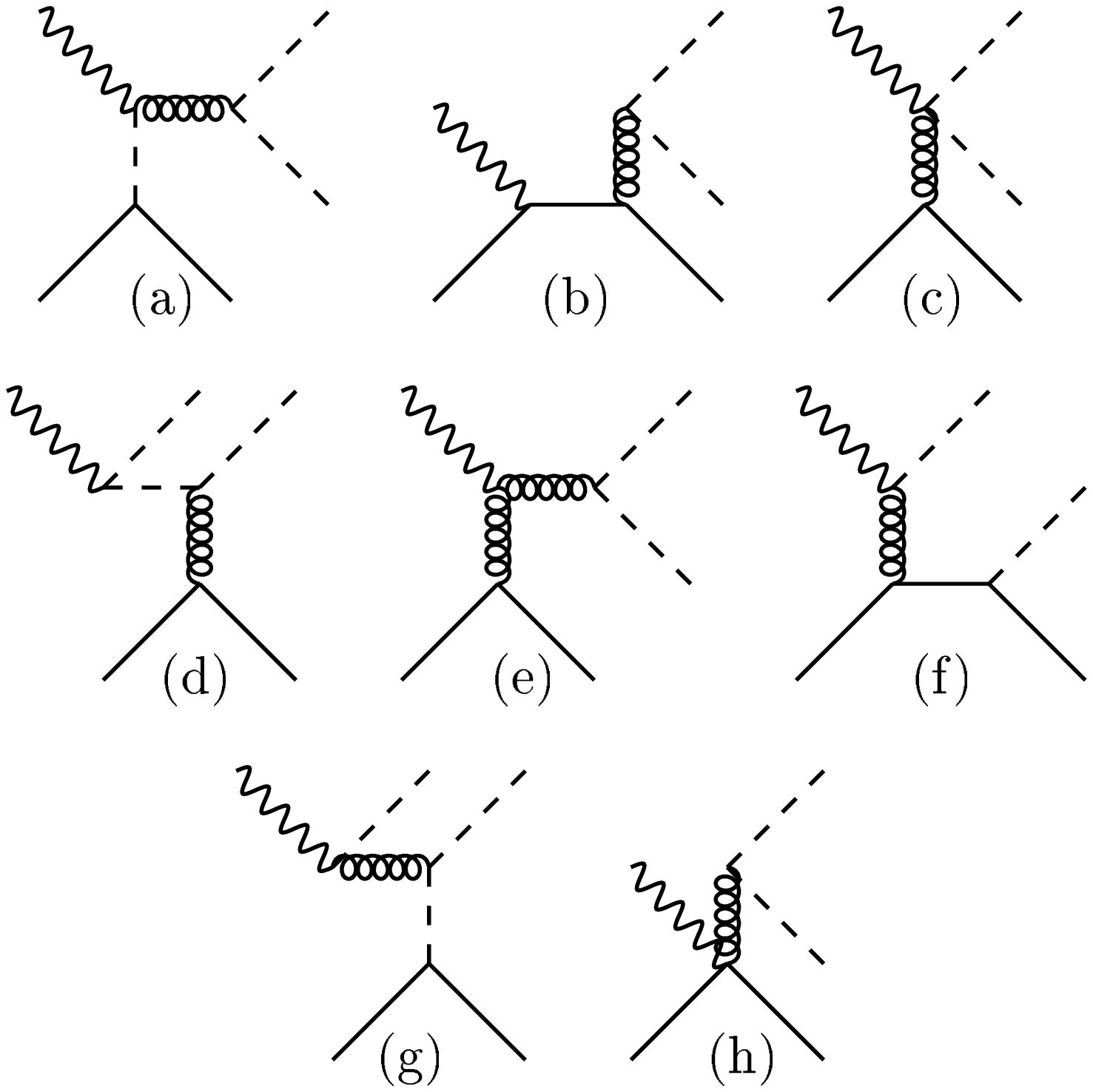}}
\vspace*{-1.5in}
\fcaption{Diagrams containing vector mesons.\label{fig:rho}}
\end{figure}

\begin{figure}
\centerline{\epsfysize=5in \epsfbox{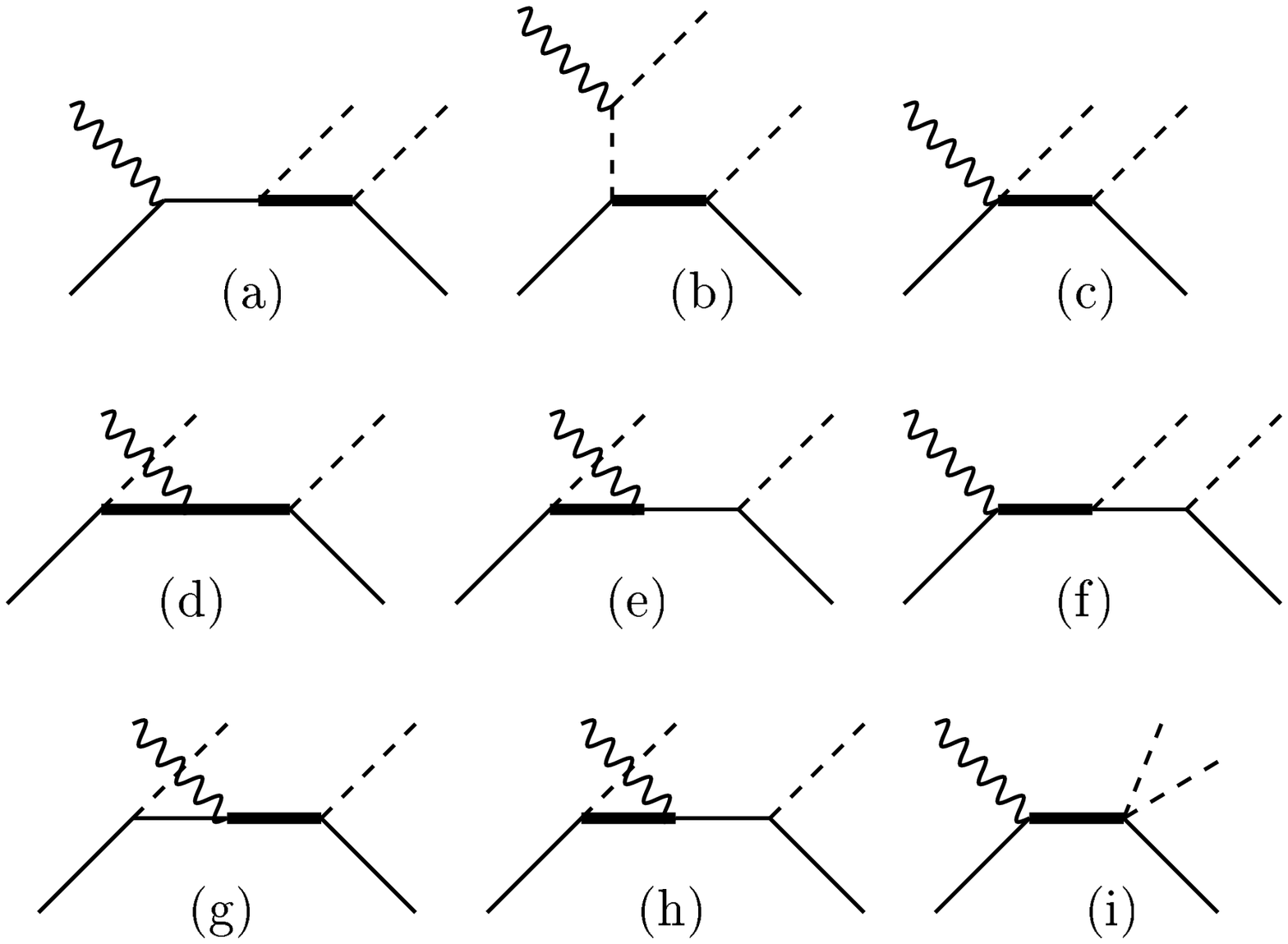}}
\vspace*{-2.2in}
\fcaption{Single-resonant diagrams.\label{fig:reson}}
\end{figure}

\begin{figure}
\centerline{\epsfysize=5in \epsfbox{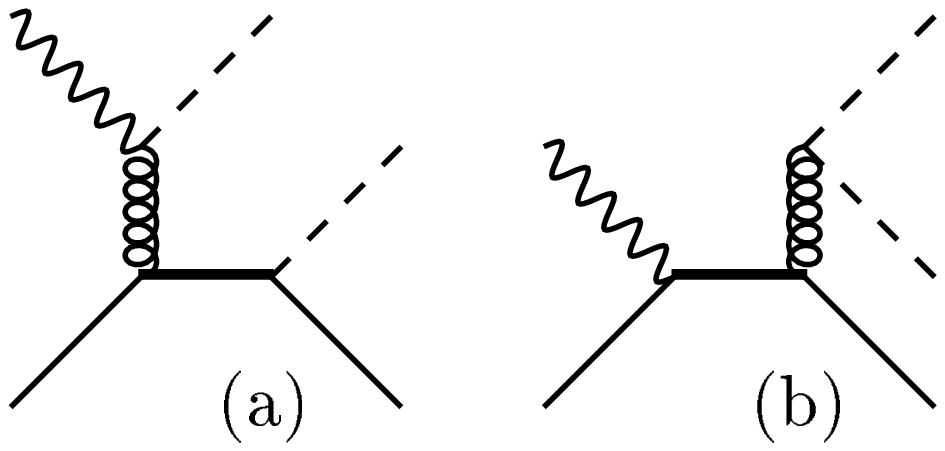}}
\vspace*{-3.9in}
\fcaption{Diagrams containing a single resonance with a vector meson.
\label{fig:resonrho}}
\end{figure}


%
\begin{figure}
\centerline{\epsfysize=5in \epsfbox{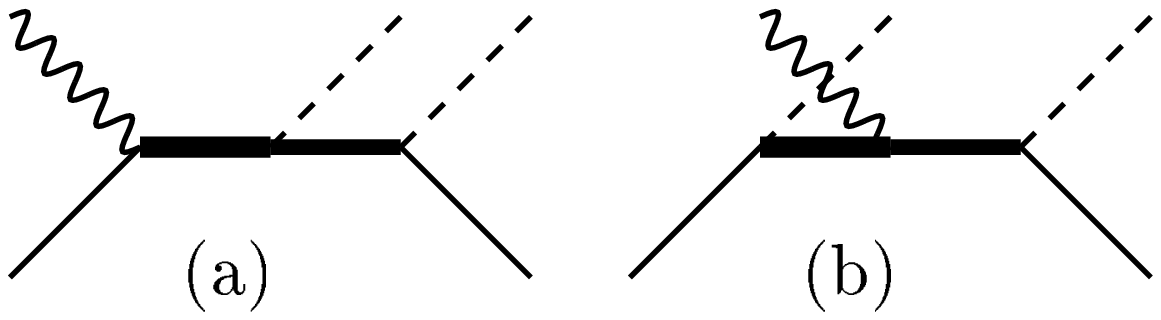}}
\vspace*{-3.9in}
\fcaption{Double-resonant diagrams.\label{fig:double}}
\end{figure}

The first set of diagrams shown in figure \ref{fig:born} are the so-called Born
terms. By this, we mean diagrams that contain neither baryon nor meson 
resonances. 
We point out that each diagram shown
actually corresponds to a set of diagrams, as we do not show all of the 
permutations that arise. For example, figure
\ref{fig:born}(a) represents 6 different diagrams, the other 5 arising from 
attaching the photon and pion lines to the nucleon line in
all possible ways. In figure \ref{fig:born}, diagram (f) arises from the 
so-called chiral anomaly
vertex, which will be the object of study of at least one experiment at 
Jefferson Lab \cite{chiral}.
Figure \ref{fig:rho} shows the diagrams that contain vector mesons in addition 
to pions and nucleons. Among these diagrams, figure
\ref{fig:rho}(a) provides an amplitude that has the same form as the chiral 
anomaly diagram of figure \ref{fig:born}. 

Figure \ref{fig:reson} shows the diagrams that contain a single baryonic 
resonance. In these diagrams, the resonance may be {\it
any} baryon resonance of any allowed spin, isospin and parity. A set of 
diagrams 
like these should therefore be included for {\it
every} resonance that one wishes to consider. Note that 
figure \ref{fig:reson}(d) 
leads to amplitudes
that are dependent on the magnetic moment(s) of the resonance. Figure 
\ref{fig:resonrho} shows the diagrams that contain both a
baryon resonance and a vector meson, and we again emphasize that a set of 
diagrams like these should be included for
every resonance that is considered. In fact, it may turn out that the 
contributions of some of these diagrams are very small.
However, it is precisely some of these small terms that must be understood if 
we are to fully comprehend the information
provided by the experimental data.

Figure \ref{fig:double} shows the diagrams that contain two different baryonic 
resonances. The diagrams of figure \ref{fig:double} depend on some `new' 
couplings, in general. For instance, figure
\ref{fig:double}(a) is proportional to the set of couplings 
$g_{N_1^*N_2^*\pi}$, 
while figure \ref{fig:double}(b) depends
on the couplings $g_{N_1^*N_2^*\gamma}$. Thus, contributions like these can 
provide new information on hadronic structure. As
with the diagrams containing a single baryon resonance, these diagrams can 
contain any combination of two baryon resonances.

\section{Conclusion}

As of this writing, we have not yet compared our calculation with experimental
results, partly because there are several questions that must be addressed
before such a comparison is made. One immediate question is that of the
validity of the effective Lagrangian approach for particles with size and
structure. We propose to accomodate the structure (and size) of these particles
by replacing the many coupling constants with form factors. 

A second, very
important question is that of unitarity in the amplitudes we calculate. This
problem could be solved by including the appropriate loops in the calculation,
but the inclusion of loop diagrams would make this calculation intractable
(there would be far too many possible diagrams for a meaningful analysis to be
done). In some sense, the inclusion of form factors at the vertices would 
partially address this problem. Perhaps a better approach would be to use these
amplitudes as input to a coupled-channel treatment.

One obvious possible criticism of our approach is the need for a very large 
number of diagrams. However, we believe that this
is inherent to the process that we are studying, and the information that we 
seek from this process. We are trying to understand
the baryon spectrum, in the hope of learning more about non-perturbative QCD. 
Thus, we must know the baryon masses and
couplings, and there are many couplings for each baryon. These are precisely 
the parameters that enter into this (or for that
matter, {\it any}) calculation of this process. Thus, comparison of this 
(or any) 
calculation with data will allow extraction of these
parameters. We note that this is essentially what is done in analyses of 
pion-nucleon 
elastic scattering data, from which we have essentially all of the information 
on the baryon spectrum that is presented in
the Particle Data Book \cite{pdg}. We note in passing that the number of 
possible 
diagrams increases dramatically if loop diagrams are
included.

One novel `result' of this calculation is the fact that the two-pion 
photoproduction process can provide new ways of testing our
understanding of hadron structure, by way of the `new' couplings 
$g_{N_1^*N_2^*\pi}$ and $g_{N_1^*N_2^*\gamma}$ discussed near the end
of the previous section. The value in this new information is obviously 
dependent on the reliability of the extraction of these numbers
from the data. By the same token, we may also be able to probe `new' aspects 
of meson structure in this process. For example, figure
{\ref{fig:rho}(e) may contain contributions from the dipole and quadrupole 
moments of the vector meson.

Clearly, cross section data alone will not be enough to delimit the possible 
choices of parameters that can provide `good' fits
to the available and expected data. There is an essential need for high quality 
polarization data. This will be crucial if we are to
exploit the many opportunities lurking in the shadows of this process. As a 
corallory we may also need to 
define new polarization observables for the two-pion process. By this, we mean 
that polarization observables must be defined for
the process $\gamma N \to N\pi\pi$, {\it not} for the assumed sub-processes 
like $\gamma N\to\Delta\pi$ and $\gamma N\to N\rho$. We
are currently examining this problem.

We note that it should be reasonably straightforward, if tedious, to extend the 
method we have presented to the
study of two-pion electroproduction. However, we make no further comments on 
this 
here, as we would like to understand as much
as possible about photoproduction before extending the model to study 
electroproduction.

We conclude by pointing out that two-pion photoproduction can provide
information related to many issues in hadron phenomenology: the question of 
missing baryons;
exotic states; the chiral anomaly; and new details of baryon and meson 
structure, to name just a few. However, the complexity of the
theoretical treatment required means that extraction of any meaningful 
information will require a Herculean effort. We believe that
what we have presented herein is one small step in that effort.

\def\pr#1#2#3{ {\it Phys. Rev.\/} {\bf#1}, (#3) #2}
\def\prl#1#2#3{ {\it Phys. Rev. Lett.\/} {\bf#1}, (#3) #2}
\def\np#1#2#3{ {\it Nucl. Phys.\/} {\bf#1}, (#3) #2}
\def\cmp#1#2#3{ {\it Comm. Math. Phys.\/} {\bf#1}, (#3) #2}
\def\pl#1#2#3{ {\it Phys. Lett.\/} {\bf#1}, (#3) #2}
\def\apj#1#2#3{ {\it Ap. J.\/} {\bf#1}, (#3) #2}
\def\aop#1#2#3{ {\it Ann. Phy.\/} {\bf#1}, (#3) #2}
\def\nc#1#2#3{ {\it Nuovo Cimento }{\bf#1}, (#3) #2}
\def\cjp#1#2#3{ {\it Can. J. Phys. }{\bf#1}, (#3) #2}
\def\zp#1#2#3{ {\it Z. Phys. }{\bf#1}, (#3) #2}


\begin{thebibliography}{9}
\bibitem{scwr} S. Capstick and W. Roberts, \pr{D47}{1994}{1993}; 
\pr{D49}{4570}{1994}.
\bibitem{luke} D. L\"uke and P. S\"oding, {\it Springer Tracts in Modern 
Physics} 
{\bf 59}
(1971) 39.
\bibitem{oset} E. Oset and J. A. Gomez-Tejedor, \np{A600}{413}{1996}.
\bibitem{mainz} A. Braghieri {\it et al.}, \pl{B363}{46}{1995}.
\bibitem{cgln} G. F. Chew, M. L. Goldberger, F. E. Low and Y. Nambu, 
\pr{106}{1345}{1957}.
\bibitem{toed} T. Oed, Rapport du Stage de D. E. A., Institut des Sciences
Nucl\'eaires, Grenoble, France, 1996. 
\bibitem{nimai} M. Benmerrouche, N. C. Mukhopadhyay and J. F. Zhang, 
\pr{D51}{3237}{1995}.
\bibitem{chiral} R. Miskimen {\it et al.}, Jefferson Laboratory proposal 
94-015.
\bibitem{pdg} Particle Data Group, R. M. Barnet {\it et al.}, \pr{D54}{1}{1996}.
\end{thebibliography}
\end{document}